%% file: NuPhys_poster_proceedings.tex
\documentclass[12pt]{article}
\usepackage{graphicx}
\usepackage{lineno}
\usepackage{amsmath}
\usepackage[font={small}]{caption}
\usepackage{subcaption}
\usepackage{url}
\usepackage{wrapfig}
\usepackage{SIunits}
\newcommand\pubnumber{}
\newcommand\pubdate{\today}

\def\napoli{Department of Physics and Astronomy\\
Universiteit Gent, Belgium}
\def\support{\footnote{for the SoLid Collaboration}}

\def\Title#1{\begin{center} {\Large #1 } \end{center}}
\def\Author#1{\begin{center}{ \sc #1} \end{center}}
\def\Address#1{\begin{center}{ \it #1} \end{center}}

\newcommand\pubblock{\rightline{\begin{tabular}{l} \pubnumber\\
         \pubdate  \end{tabular}}}
\newenvironment{Abstract}{\begin{quotation}  }{\end{quotation}}
\newenvironment{Presented}{\begin{quotation} \begin{center} 
             PRESENTED AT\end{center}\bigskip 
      \begin{center}\begin{large}}{\end{large}\end{center} \end{quotation}}

\input econfmacros.tex

\begin{document}
\begin{titlepage}
\pubblock

\vfill
\Title{SoLid: Search for Oscillation with a $^6$Li Detector at the BR2 research reactor}
\vfill
\Author{ Ianthe Michiels\support}
\Address{\napoli}
\vfill
\begin{Abstract}
An introduction to the SoLid detection principle and some of the first results obtained with the large scale test module SM1.
\end{Abstract}
\vfill
\begin{Presented}
 NuPhys2015, Prospects in Neutrino Physics\\
Barbican Centre, London, UK,  December 16--18, 2015
\end{Presented}
\vfill
\end{titlepage}
\def\thefootnote{\fnsymbol{footnote}}
\setcounter{footnote}{0}

\section{Introduction}
Emphasized by this year's Nobel Prize in Physics, it can safely be said that neutrino oscillations are a hot topic in the world of particle physics. The experimental proof that neutrinos can transform into each other was first evidence of physics beyond the Standard Model~\cite{superK}. More recent experiments have shown that there could be more surprises in the neutrino sector, like the existence of a fourth, sterile neutrino. This has been suggested by the results of nuclear reactor neutrino experiments, where the flux of antineutrinos coming from the reactor shows a deficit at short reactor-detector distances, when compared to theoretical calculations~\cite{Mention}. The appearance of the effect at short baselines implies the hypothetical sterile neutrino to be \emph{light}, with oscillation parameters around $\Delta m^2 \sim 1$ eV$^2$ and $\sin^2(2\theta) \sim 0.1$~\cite{Abazajian}. One of the experiments designed to investigate the hypothesis of this sterile neutrino is the SoLid experiment. It uses the compact BR2 research reactor from the SCK$\bullet$CEN in Mol, Belgium, to perform reactor antineutrino flux measurements at very short baseline.

\section{The SoLid experiment}
\subsection{BR2 reactor site}
\begin{figure}[h!]
\centering
\includegraphics[height=2.11in]{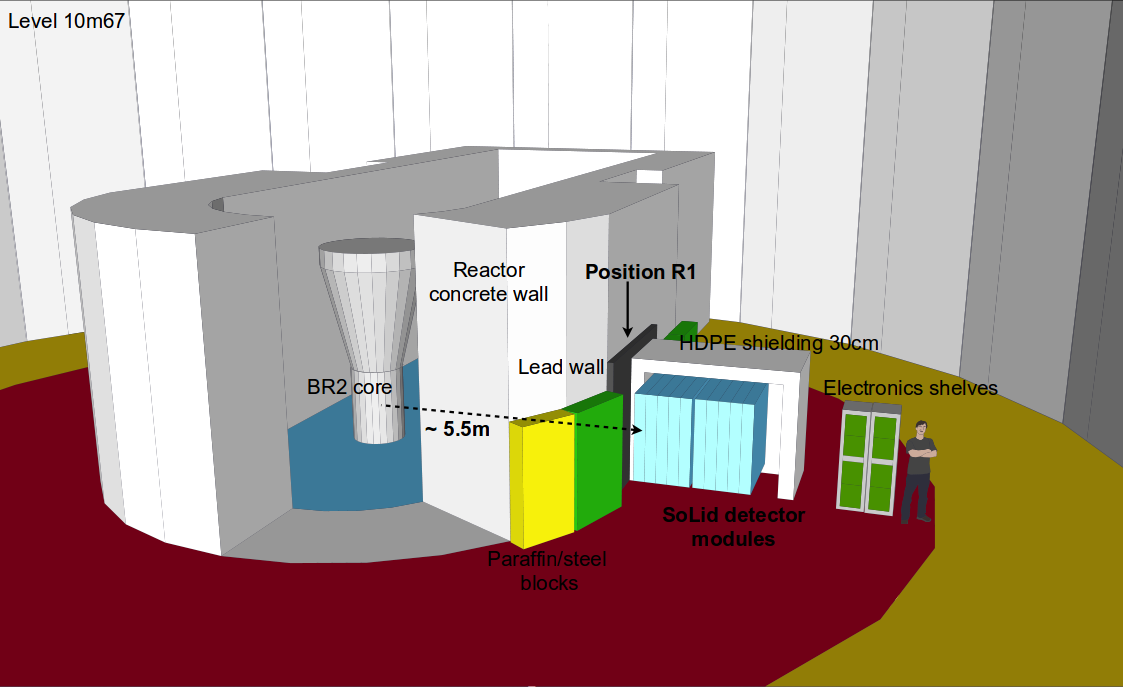}
\caption{The SoLid experiment in the BR2 reactor hall}
\label{fig:br2}
\end{figure}

Figure~\ref{fig:br2} illustrates the set-up of the SoLid experiment in the BR2 reactor hall. 
BR2 is a tank-in-pool research reactor that can reach a power of up to 100 MW. Its very compact core has a diameter of 1.1 m, which implies that it can serve as an almost pointlike high intensity source of antineutrinos, plus it allows the SoLid experiment to take measurements at baselines as short as 5.5 m.
Since currently no other experiments take place at the same floor of the reactor building, all other experiment gates in the hall are closed, which provides relatively stable background conditions for this site.

\subsection{Detection principle}
\begin{figure}[h!]
\centering
\begin{minipage}[b]{0.45\textwidth}
\includegraphics[height=2in]{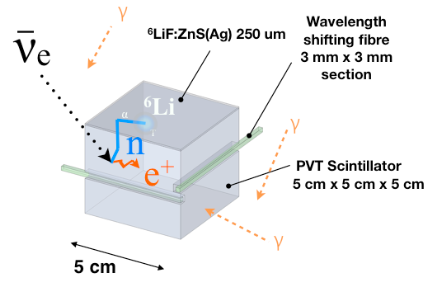}
\caption{The interaction of an electron antineutrino with a SoLid detector cube. The light pulses created in the cube are transported towards read-out electronics by a pair of wavelength shifting fibres.}
\label{fig:cubeInt}
\end{minipage}
\hspace{9mm}
\begin{minipage}[b]{0.47\textwidth}
\includegraphics[height=2.2in]{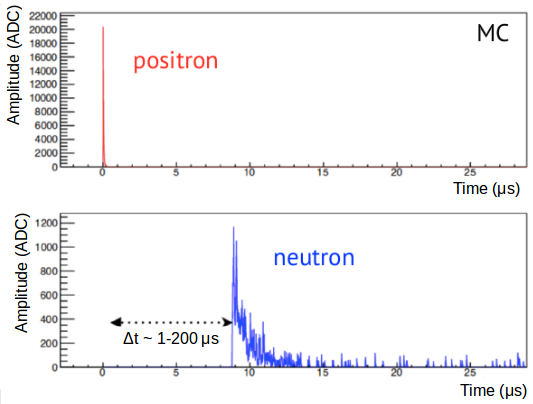}
\caption{The signature of an IBD interaction in the SoLid detector. The thermalization of the neutron creates a typical time delay $\Delta t$ of  $\mathcal{O}(100)\ \mu$s.}
\label{fig:ibdSig}
\end{minipage}

\end{figure}

The detection of the reactor electron antineutrinos is based on an inverse beta decay (IBD) reaction:
\begin{equation}
\bar{\nu}_e + p \rightarrow e^+ + n.
\end{equation}
SoLid applies a newly developed detector technology using ($5 \times 5 \times 5$) cm cubes of polyvinyl toluene (PVT) scintillator combined with $^6$LiF:ZnS(Ag) sheets to capture respectively the positron and neutron from such an IBD interaction (cf. figure~\ref{fig:cubeInt}). 

The charged positron is almost immediately - within $\mathcal{O}(10)$ ns - absorbed by the PVT, which results in a short, intense light pulse. The neutron first thermalizes before it is captured by $^6$Li:
\begin{equation}
n + {^6}\text{Li} \rightarrow {^3}\text{H} + \alpha + 4.78\ \text{MeV}.
\end{equation}
The energy released in this reaction excites the ZnS(Ag) and the subsequent decays of these excited states create detectable light pulses. A more detailed description of the detection principle can be found in~\cite{Ryder}. 

Figure~\ref{fig:ibdSig} gives an illustration of the expected IBD signal, from which one can see that identification of the IBD reaction products can be based on pulse shape discrimination. The reconstruction of a full IBD event relies on the specific time delay $\Delta t$ between the positron and neutron absorption processes and their small spatial separation.

A SoLid detector module exists out of a stack of these small PVT cubes, and is thus highly segmented, which enables precise localization of interactions. This grants the opportunity to apply spatial cuts for background reduction and adds the possibility to perform direction reconstruction. As a consequence, the SoLid experiment is very robust in discriminating signal from background.

\subsection{SubModule 1}
\begin{figure}[htb]
\centering
\begin{minipage}[b]{0.38\textwidth}
\center
\includegraphics[trim={0 1.2cm 0 4mm}, clip, height=2in]{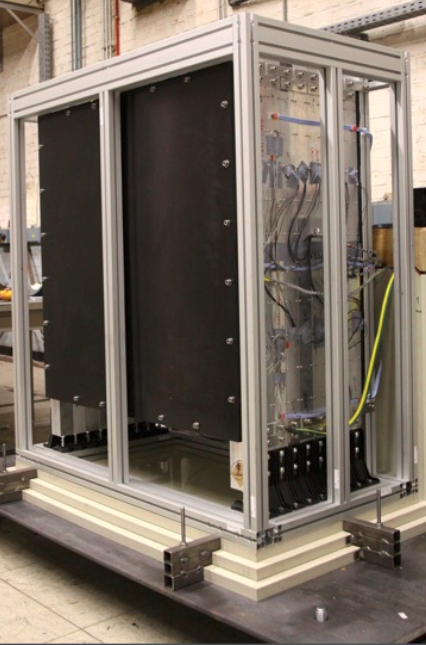}
\caption{The first large scale detector module of the SoLid experiment; SM1.}
\label{fig:SM1}
\end{minipage}
\hspace{4mm}
\begin{minipage}[b]{0.52\textwidth}
\center
\includegraphics[height=1.7in]{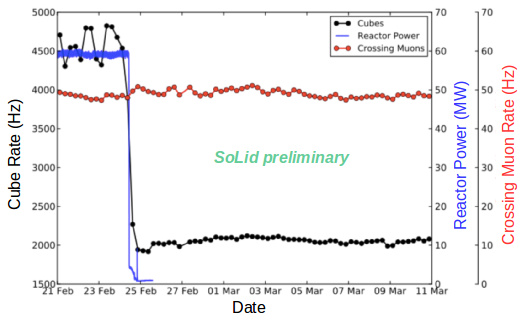}
\caption{The period of data taking with SM1, illustrated by the crossing muon rate and the rate of detected signals per cube, the latter of which is clearly correlated with the reactor power.}
\label{fig:dataPeriod}
\end{minipage}
\end{figure}

The SoLid Collaboration has succesfully commissioned a first large scale submodule (SM1) of the detector in the winter of 2014/2015. This module is built out of 2304 PVT cubes that are distributed over 9 detector planes of $16 \times 16$ cubes, resulting in a total weight of 288 kg (cf. figure~\ref{fig:SM1}). See~\cite{Moortgat} for more information.

From December 2014 to March 2015 the SM1 detector recorded 3 to 4 days of reactor on data, and approximately 1 month of reactor off data (cf. figure~\ref{fig:dataPeriod}). Afterwards, some additional calibration runs were performed with radioactive sources; $^{60}$Co and AmBe source calibrations took place in April 2015 and a $^{252}$Cf in situ measurement was taken in August 2015.

\section{Background analysis with SM1 data}

\begin{figure}[htb]
\centering
\begin{minipage}[b]{0.47\textwidth}
\includegraphics[height=2.1in]{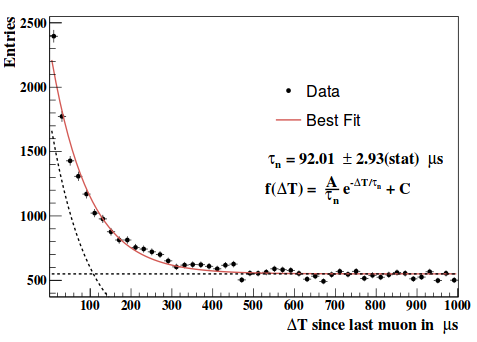}
\caption{The time difference $\Delta T$ between a muon and a neutron-like signal shows an exponential distribtution (with flat background) with a decay constant $\tau_n = (92.01 \pm 2.93)\ \mu$s.}
\label{fig:muon}
\end{minipage}
\hspace{4mm}
\begin{minipage}[b]{0.47\textwidth}
\centering
\includegraphics[height=2.15in]{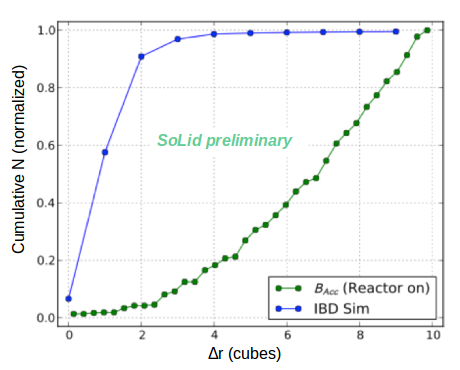}
\caption{The cumulative distribution of the distance $\Delta r$ between the prompt and delayed signal (in units of 5 cm cubes) for accidental background (green) and simulated IBD events (blue).}
\label{fig:deltaR}
\end{minipage}
\end{figure}

Because the SoLid experiment is conducted at sea-level, an important background component is induced by cosmic muons. These highly energetic particles create spallation neutrons in or close to the detector and consequently contribute to the correlated backgrounds of the experiment.\footnote{In this text, two types of backgrounds are distinguished. \emph{Correlated backgrounds} refer to consecutive signals with a specific timing relation. This in contrast with \emph{accidental backgrounds}, which originate from random coincidences of uncorrelated signals.} Therefore, the muon induced backgrounds are analyzed for a better understanding of the background itself and of the detector response~\cite{Kalousis}.
Figure~\ref{fig:muon} shows, for example, how the SM1 data was used to determine a capture time $\tau_n$ for muon induced neutrons.

The fact that the SoLid experiment is located in the close proximity of a nuclear reactor brings
along some additional types of backgrounds. Examples of these are environmental neutrons and reactor
gamma rays, of which the latter generate a high level of accidental background, since their signature
can hardly be discriminated from an IBD positron interaction. A random coincidence of an environmental neutron and a reactor gamma ray can therefore mimic an IBD event.

As mentioned above, the SoLid detector is a highly segmented volume which provides the experiment with the ability to apply strong radial cuts. Figure~\ref{fig:deltaR} shows the cumulative distribution of the distance in number of cubes between the detected neutron and the prompt electromagnetic signal for both accidental background and simulated IBD events. The distinct behaviour of the two curves shows that a radial cut can be very effective in reducing the accidental background of an IBD search.

By applying other topological restrictions and energy cuts, the accidental background can be reduced to very low levels~\cite{Moriond}. Other analyses concerning the reduction of correlated backgrounds and the selection of antineutrino candidates, using the SM1 data, are currently ongoing.

\section{Summary}

The SoLid Collaboration has succesfully commissioned a first large scale submodule (SM1) of the detector in the winter of 2014/2015. SM1 demonstrated the large scale use of the detector technology and provided the collaboration with valuable data to tune the electronics and analysis software tools. A large part of the SM1 data analyses focussed on techniques to understand the various types of background, present at the BR2 reactor site. The gathered insights will benefit the construction and commissioning of the full scale (1-2 tonne) detector in the second half of 2016. The SoLid experiment will thus continue to probe the light sterile neutrino oscillation region, aiming for maximum sensitivity in 2 years of data taking.

\end{document}

%% file: econfmacros.tex
\def\beq{\begin{equation}}
\def\eeq#1{\label{#1}\end{equation}}
\def\eeqn{\end{equation}}

\def\beqa{\begin{eqnarray}}
\def\eeqa#1{\label{#1}\end{eqnarray}}
\def\eeqan{\end{eqnarray}}

\let\bar=\overbar

\def\Dslash{\not{\hbox{\kern-4pt $D$}}}
\def\dslash{\not{\hbox{\kern-2pt $\del$}}}

\def\msb{{\bar{\ssstyle M \kern -1pt S}}}

